\documentclass[10pt,a4paper,fleqn,accepted=2019-10-02]{quantumarticle}
\pdfoutput=1
\usepackage{CJK}
\usepackage[utf8]{inputenc}
\usepackage[english]{babel}
\usepackage[utf8]{inputenc}
\usepackage{placeins}
\usepackage[T1]{fontenc}
\usepackage{amsmath}
\usepackage{amsfonts}
\usepackage{amssymb}
\usepackage{graphicx}
\usepackage[left=2cm,right=2cm,top=2cm,bottom=2cm]{geometry}
\usepackage[version=3]{mhchem}
\usepackage{appendix}
\usepackage{multirow}
\usepackage{array}
\usepackage{siunitx}
\usepackage{subcaption}
\usepackage[font=footnotesize,margin=20pt,labelfont=bf,format=plain]{caption}
\usepackage{chngcntr}
\usepackage{wrapfig}
\usepackage{url}
\usepackage[version=3]{mhchem}
\usepackage{braket}
\usepackage{textcomp}
\usepackage{braket}
\usepackage{authblk}
\usepackage{multicol}
\usepackage{bm}
\usepackage{hyperref}
\usepackage{cite}
\usepackage{cleveref}

\begin{document}

\title{Pilot-assisted intradyne reception for high-speed continuous-variable quantum key distribution with true local oscillator}
\date{}

\author[1,2]{Fabian Laudenbach}
\email{fabian.laudenbach@ait.ac.at}
\author[1]{Bernhard Schrenk}
\author[1]{Christoph Pacher}
\author[1]{Michael Hentschel}
\author[3]{Chi-Hang Fred Fung}
\author[3]{Fotini Karinou}
\author[3]{Andreas Poppe}
\author[3]{Momtchil Peev}
\author[1]{Hannes H\"{u}bel}

\affil[1]{Security \& Communication Technologies, Center for Digital Safety \& Security,\protect \\AIT Austrian Institute of Technology GmbH, Giefinggasse 4, 1210 Vienna, Austria}
\affil[2]{Quantum Optics, Quantum Nanophysics \& Quantum Information, Faculty of Physics,\protect \\University of Vienna, Boltzmanngasse 5, 1090 Vienna, Austria}
\affil[3]{Optical and Quantum Laboratory, Munich Research Center,\protect \\Huawei Technologies Duesseldorf GmbH, Riesstrasse 25-C3, 80992 Munich, Germany \protect \\ \quad}

\maketitle

\begin{abstract}
We present a pilot-assisted coherent intradyne reception methodology for CV-QKD with true local oscillator. An optically phase-locked reference tone, prepared using carrier-suppressed optical single-sideband modulation, is multiplexed in polarisation and frequency to the 250 Mbaud quantum signal in order to provide optical frequency- and phase matching between quantum signal and local oscillator. Our concept allows for high symbol rates and can be operated at an extremely low excess-noise level, as validated by experimental measurements.
\end{abstract}

\section{Introduction}

Quantum key distribution using continuous variables (CV-QKD)~\cite{grosshans2002continuous, grosshans2003virtual, scarani2009security, weedbrook2012continuous, laudenbach2017continuous} is currently regarded as one of the main contenders for a full-scale deployment of quantum cryptography. Its advantages over traditional qubit-based implementations include higher key rates and, more importantly, the ability to use established telecom technology (I/Q-modulation, Mach-Zehnder pulse carving, coherent detection) rather than complex and costly components required for discrete-variable QKD, in particular single-photon detectors. Unlike counting polarisation- or phase-encoded single photons, in CV-QKD the raw key is established by encoding the quadrature components of weak coherent states $\ket{\alpha}=\ket{I_{A}+iQ_{A}}$. The information read-out is performed by coherent detection where the weak quantum signal is mixed with an optically strong reference laser, the so-called local oscillator (LO), at a balanced beamsplitter. The difference in optical power at the output ports of the beamsplitter is then proportional to the quadrature $I$ or $Q$, depending on the phase $\Theta$ of the LO: $\Delta P \propto |\alpha_{\text{LO}}|  \left( I \cos{\Theta}+Q\sin{\Theta} \right)$. The PIN diodes used for the power measurement can operate at high rates (up to $\sim \SI{10}{GHz}$) and are at the same time highly efficient and low-priced. This compares beneficially to avalanche photo diodes used in discrete-variable QKD which are both limited in the detection rate (by their dead time after a counting event) and quantum efficiency but are at the same time several times as expensive. Moreover, the facilitated integration of balanced detectors onto photonic chips crucially supports the miniaturisation of CV-QKD receivers for ubiquitous quantum-information applications.

For coherent detection the two lasers, signal transmitter and LO, can either have the same optical frequency ($f_{T} = f_{\text{LO}}$, referred to as \emph{homodyne} detection), or a slight frequency offset smaller than the symbol rate ($|f_{T} - f_{\text{LO}}|<R_{\text{sym}}$, \emph{intradyne} detection) or even a greater frequency offset where the data is downconverted to an intermediate frequency $|f_{T} - f_{\text{LO}}|>R_{\text{sym}}$ in the RF spectrum (\emph{heterodyne} detection).\footnote{Note the different terminology compared to most CV-QKD literature where homodyne and heterodyne detection refers to measurement in one or two quadrature bases, respectively.}  However, no matter which of the above schemes is implemented, coherent detection requires for the signal laser and the LO to retain a stable and well-known frequency- and phase relation. One natural and simple way to provide for this requirement is to have the signal and the LO originate from the same laser source, as it was implemented in early realisations of CV-QKD~\cite{lodewyck2007quantum, qi2007experimental, fossier2009field, jouguet2013experimental}. This (as we call it `in-line local-oscillator') approach, however, requires for the LO to be jointly transmitted with the quantum signal which not only severely limits the total LO power available at the receiver (due to channel loss), but also disturbs the quantum- (and other DWDM) channels in the fibre. More importantly even, it opens grave security loopholes due to possible side-channel attacks that an eavesdropper can perform on the LO~\cite{haeseler2008testing, huang2013quantum, ma2013wavelength, qin2013saturation, jouguet2013preventing, ma2013local}.

In order to provide for security of the key-exchange procedure and compatibility with telecom fibre infrastructure, it is therefore unavoidable to generate the LO locally at the receiver (`\emph{true} local oscillator', or sometimes `\emph{local} local oscillator', LLO). Since in this scheme the signal and LO laser are mutually independent, the LLO scheme requires a phase- (and frequency) synchronisation to allow for coherent detection. As already widely established in classical communication, carrier-phase recovery does not necessarily require to adjust and lock the relative phase and frequency between the two lasers \emph{ahead} of measurement. Instead, the measurement can be performed with an arbitrary relative phase, yielding a corresponding rotation of Bob's phase-space coordinates with respect to Alice's. If Bob has knowledge about the phase- and frequency difference between the two lasers, he can counter-rotate his coordinate axes \emph{post-measurement} and reconcile his data with Alice's reference frame. In standard quadrature-amplitude modulation (QAM) used for telecommunication, the phase- and frequency correction is directly extracted from the data signal. Since in CV-QKD the quantum signal itself is too weak to allow for a precise phase- and frequency measurement, Alice will prepare a strong second signal, represented by a fixed and well-known point in phase space. Originating from the same laser as the quantum states, this reference signal carries all the frequency and phase information that Bob needs in order to estimate the phase- and frequency difference between his laser and Alice's. He will mix his local oscillator with the quantum and reference signal and measure the quadratures of both independently. The measurement of the reference signal allows him to monitor the phase drift over time and to apply the reverse rotation to the individual quantum measurements accordingly.

In first demonstrations of the above method a time-multiplexed scheme was adopted, where strong reference pulses were temporally interleaved with the quantum signal~\cite{huang2015high, qi2015generating, soh2015self}, in some cases enhanced by polarisation-multiplexing~\cite{wang2018pilot, wang2018high}. Although straightforward to implement, this method comes with certain impairments. Firstly, in the time-multiplexing scheme the additional synchronization pulses will reduce the rate of the quantum signal since Alice needs to reserve periodic time slots in her pulse train for the reference signal. Secondly, the quadratures of the quantum signal are not measured at exactly the same time as the synchronisation quadratures, i.e. phase changes which are fast compared to the symbol rate will not be compensated. Finally, if the quantum- and the reference signal are measured with the same balanced receivers, the allowed optical power of the reference pulse is restricted by the saturation limit of the PIN diodes which is usually very low for low-noise receivers as required for CV-QKD.\footnote{For instance, the optical-power damage threshold for the low-noise detectors used in our experiment (\emph{Insight BPD-1}, noise-equivalent power $< \SI{5}{pW/\sqrt{Hz}}$) amounts to $\SI{10}{dBm}$.} On the other hand, routing of signal- and reference pulses to designated receivers respectively (low-noise for quantum, high-saturation limit for the reference) requires cumbersome and fast switching, when the two are multiplexed in the time degree of freedom only.

In addition to the sequential transmission of signal- and reference pulses, Ref.~\cite{marie2017self} proposes a scheme based on modulation displacement, where each symbol encoded by Alice is added to a fixed and well-known offset in amplitude $|\Delta|$ and phase $\theta_{\Delta}$. Therefore the transmitted coherent states are represented as $\ket{\alpha}=\ket{I_{A} |\Delta| \cos \theta_{\Delta} + Q_{A} |\Delta| \sin \theta_{\Delta}} \equiv \ket{\alpha_{A} + \Delta}$ where $\alpha_{A}$ is a weak coherent state carrying the quantum information and the offset $\Delta$ is a stronger coherent state carrying the phase reference. Bob performs simultaneous measurements on the states in the $I$- and $Q$-basis (in CV-QKD literature often referred to as `heterodyne' measurement) and extracts the phase estimation from the displacement of Alice's modulation. Unlike the methods based on time-multiplexing, in this approach the accuracy of the phase estimation is not independent from the modulation variance (since a higher amplitude of the quantum signal will, paradoxically, reduce the SNR of the offset-phase measurement). Moreover, this scheme suffers from the trade-off that the balanced receivers have to operate at low electronic noise (required for CV-QKD security) but at the same time need to have a sufficiently high saturation limit in order to allow for an accurate measurement of the strong offset state.

Inspired by the practices of the telecom industry, multiplexing by the modulation frequencies of the quantum signal and a reference signal, a so-called pilot tone, has been explored as a promising alternative to sequential modulation of reference pulses. In the simplest form of frequency-multiplexing, the frequency-upconverted quantum signal and pilot tone are of the same polarisation and detected using one and the same balanced heterodyne receiver~\cite{kleis2015simple,brunner2017low, comandar2017flexible, kleis2017continuous}. Starting in 2016, we have been demonstrating that multiplexing of the quantum signal and a pilot tone in the frequency- and polarisation domain~\cite{schrenk2016pilot, schrenk2017high, laudenbach2017pilot} allows for improved suppression of cross-talk and for dedicated pilot and quantum receivers.

In this work we present a complete pilot-assisted coherent \emph{intradyne} reception methodology (frequency offset between transmitter laser and LO much smaller than the symbol rate, baseband detection of quantum data) in which an optically phase-locked reference tone is multiplexed to the actual quantum signal in both, modulation frequency \emph{and} polarisation. The quantum signal was generated using quadrature-phaseshift-keying (QPSK) at a symbol rate of $R_{\text{sym}}=\SI{250}{Mbaud}$; the pilot tone was generated using single-sideband modulation with optical-carrier suppression (oCS-SSB; more details in Sec.~\ref{sec_trans}) at $f_{P}=\SI{1}{GHz}$. As opposed to Ref.~\cite{kleis2017continuous}, the exploitation of the polarisation degree of freedom allows us to (1) efficiently avoid crosstalk of the strong pilot to the weak quantum signal, (2) perform adequate power-levelling of signal and pilot as well as (3) the use of optimised receivers, accounting for the particular requirements of the weak quantum signal and strong pilot tone, respectively. Moreover, our intradyne-reception architecture can be operated at a much lower sampling rate compared to schemes based on heterodyne detection.

\section{Pilot-Assisted Continuous-Variable Detection Scheme with Local Oscillator at Receiver}

\subsection{Transmitter Setup} \label{sec_trans}

\begin{figure*}
\centering
\includegraphics[width=0.9\linewidth]{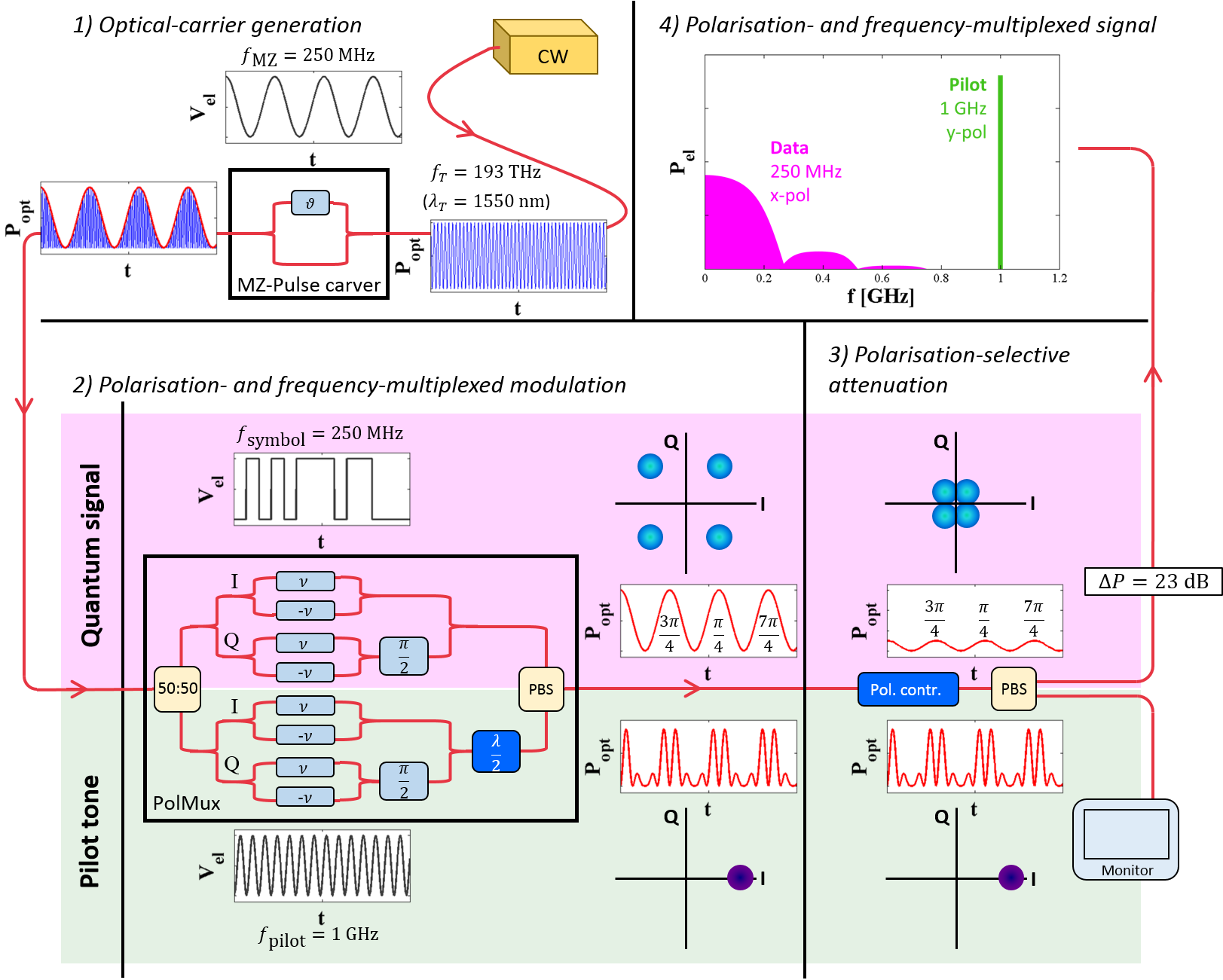}
\caption{Schematics of the transmitter setup. A $\SI{250}{MHz}$ pulse train, carved by Mach-Zehnder intensity modulation of a $\SI{1550}{nm}$ CW laser, is fed into the polarisation-multiplexed I/Q modulator (PolMux). The PolMux branches the pulses into two halves where the quantum signal and pilot tone are I/Q modulated independently (light-blue boxes representing phase rotations and the dark-blue box representing a polarisation rotation by a $\SI{90}{\degree}$ half-wave plate). The quantum branch is modulated with a $\SI{250}{Mbaud}$ QPSK pattern, driven by a pseudorandom binary sequence (PRBS7). For the pilot tone, we performed optical single-sideband modulation with suppressed carrier, driven by a $\SI{1}{GHz}$ cosine function (see Fig.~\ref{spectra}(a)). After recombination of the two branches under orthogonal polarisation, the quantum signal was attenuated (see Fig.~\ref{spectra}(b)) to adhere to the security requirements of CV-QKD while at the same time retaining a strong pilot amplitude to allow for a high SNR, as required for accurate phase recovery.}
\label{polmux}
\end{figure*}

\begin{figure*}
\centering
\subcaptionbox{}
	[0.45\linewidth]{\includegraphics[width=0.44\textwidth]{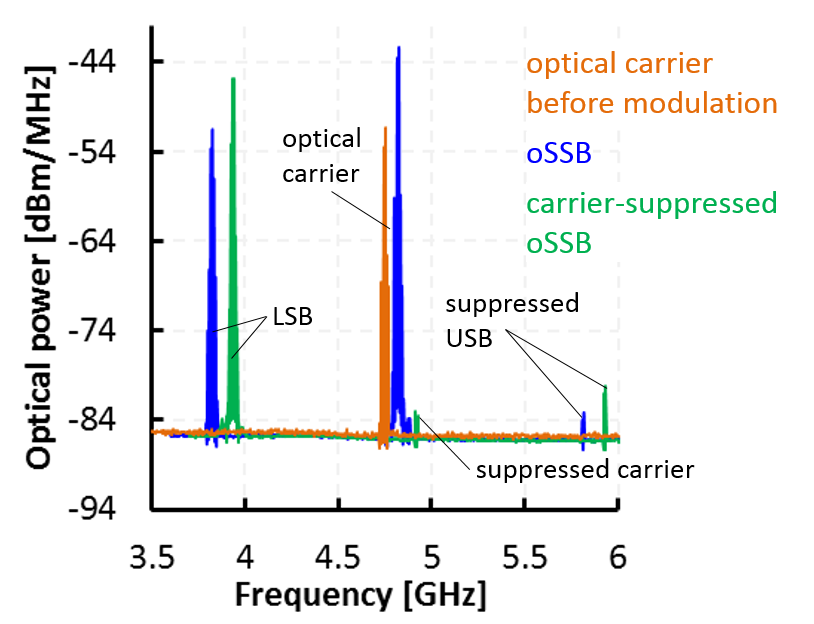}}
\subcaptionbox{}
	[0.45\linewidth]{\includegraphics[width=0.44\textwidth]{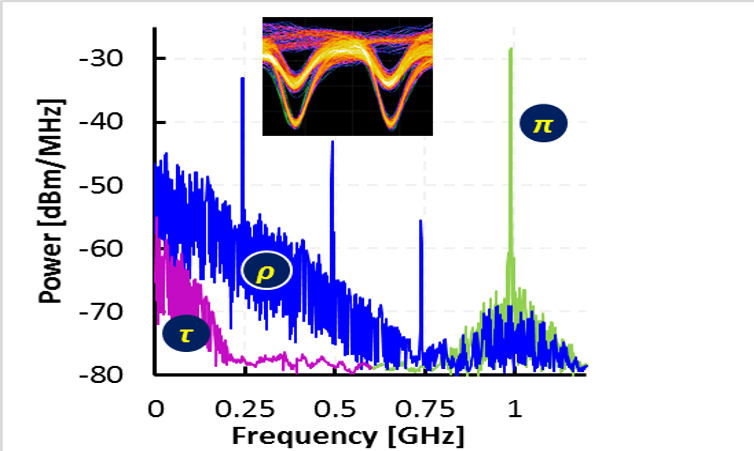}}
\caption{(a) Optical single-sideband pilot tone (mixed to an intermediate frequency of $\SI{5.5}{GHz}$) without (blue) and with (green) optical carrier suppresion. LSB (USB) indicates the lower (upper) sideband. (In the plot, the traces were shifted $\SI{0.1}{GHz}$ relative to one another along the $x$-axis to improve the readability of the graph.) (b) Signal spectra after polarisation-selective attenuation. Using a fibre-based polarisation controller and polarising beamsplitter with finite extinction ratio, we suppressed the pilot tone at the monitored output port of a PBS~($\rho$). Consequently, the transmitted output was fed with a strong pilot~($\pi$) and weak quantum signal~($\tau$) in orthogonal polarisation, the power difference amounting to $\sim \SI{23}{dB}$. The peaks in the blue trace indicate the harmonics of the symbol rate.}
\label{spectra}
\end{figure*}

The experimental transmitter setup is illustrated in Fig.~\ref{polmux}. The optical carrier at $\lambda_{T}= \SI{1550.12}{nm}$ with a linewidth of $\SI{400}{kHz}$ (\emph{Teraxion PS-LM}) was amplitude-modulated from continuous wave to $\SI{250}{MHz}$ pulses using a Mach-Zehnder pulse carver (\emph{Optilab IML-1550-40-PM}). The pulse-carving (1) facilitates the time synchronisation with the receiver (2) allows for interleaving multiple signals by time-division multiplexing and (3) removes a potential side channel that Eve may use to gain information by monitoring the transition between two symbols.

The pulse train was fed into a polarisation-multiplexing (`PolMux') IQ modulator (\emph{Fujitsu FTM1977HQA}) which allows for independent modulation of the orthogonally polarised pilot tone and quantum signal, preserving their locked frequency and phase relation. The quantum- and pilot modulators were controlled by an arbitrary-waveform generator (AWG, \emph{Keysight M8195A}). The quantum tributary was phase modulated (modulation index 0.94) to yield a QPSK signal according to a repeating pseudorandom binary sequence of length $2^{7}-1$ (PRBS7) at a symbol rate of $R_{\text{sym}} = \SI{250}{Mbaud}$.

The pilot tone was modulated with a $f_{P} = \SI{1}{GHz}$ cosine function (modulation index 0.76), representing a fixed symbol, static in phase space. For preparation of the pilot we performed single-sideband modulation with suppressed optical carrier (oCS-SSB) which is motivated as follows: Modulation of an optical signal carrier with a cosine function will create two sidebands, spaced at modulation frequency $f_{\text{mod}}$ (in our case $\SI{1}{GHz}$) below and above the optical carrier frequency $f_{T}$ (in our case $\approx \SI{193}{THz}$). At coherent detection these sidebands are downconverted to the frequencies $|f_{T}-f_{\text{LO}} \pm f_{\text{mod}}|$. In case of homodyne ($f_{T}=f_{\text{LO}}$) or intradyne ($f_{T}\approx f_{\text{LO}}$) reception, both sidebands will be projected to (almost) the same RF frequency, causing harmful self-interference. Since both sidebands carry the exact same information it is sufficient to transmit only one of them while the other one is suppressed by appropriate phase settings of the Mach-Zehnder I/Q modulator (Fig.~\ref{spectra}(a), blue curve). On top of SSB modulation, suppression of the optical carrier at $f_{T}$ (Fig.~\ref{spectra}(a), green curve) helps to avoid a detrimental beat note in the low-frequency regime $|f_{T}-f_{\text{LO}}|$ that would interfere with the quantum signal. This effect should in principle be suppressed for cross-polarised quantum signal and pilot tone but does in practice still occur due to imperfect polarisation splitting by the optical hybrid (in our case \emph{Kylia COH28-X} with a polarisation splitting ratio of $\SI{20}{dB}$). The striking advantages of optical carrier suppression in terms of excess noise are verified by our experimental results, discussed in Sec.~\ref{sec_perf}.

In order to ensure security of the protocol, the quantum signal is supposed to be sufficiently weak (depending on the channel length and noise level $\sim 0.1\text{--}10$ photons per symbol). On the other hand, the pilot tone is required to be as strong as possible to allow for an accurate phase measurement. Therefore, we performed a polarisation-dependent attenuation, reducing the optical power of the quantum signal by $-\SI{23}{dB}$ with respect to the pilot tone. This power-levelling between pilot and data signal was performed while preserving optical phase-locking and facilitated through selective attenuation on the polarisation tributaries using a fibre-based polarisation controller (PC) and polarising beamsplitter (PBS). By alignment of the PC, the optical pilot power at $\SI{1}{GHz}$ is suppressed in the monitored port of the PBS ($\rho$ in Fig.~\ref{spectra}(b)) and therefore maximised in the output port ($\pi$). Taking advantage of the finite PBS extinction ratio ($\SI{23}{dB}$), a strongly attenuated fraction of the quantum signal ($\tau$) gets transmitted to the output port along with the strong pilot. The inset in Fig.~\ref{spectra}(b) shows the eye diagram of the QPSK quantum signal with the characteristic dips.

\subsection{Receiver Setup}

\begin{figure*}
\centering
\subcaptionbox{}
	[0.9\linewidth]{\includegraphics[width=0.9\textwidth]{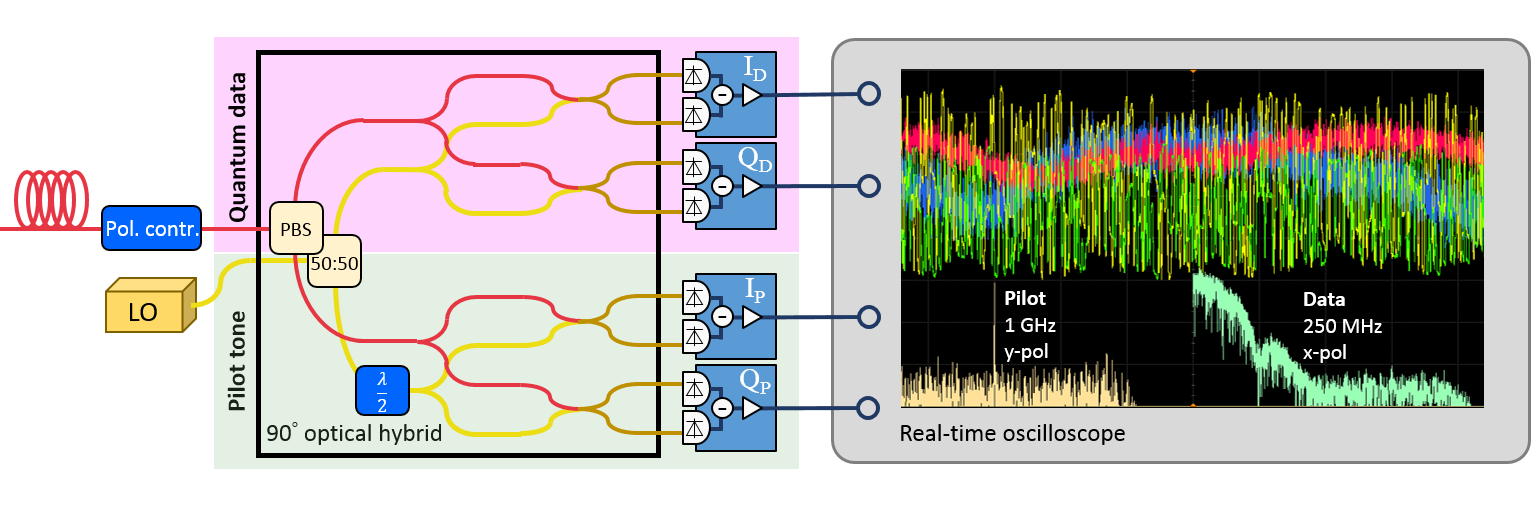}}
\subcaptionbox{}
	[0.39\linewidth]{\includegraphics[width=0.35\textwidth]{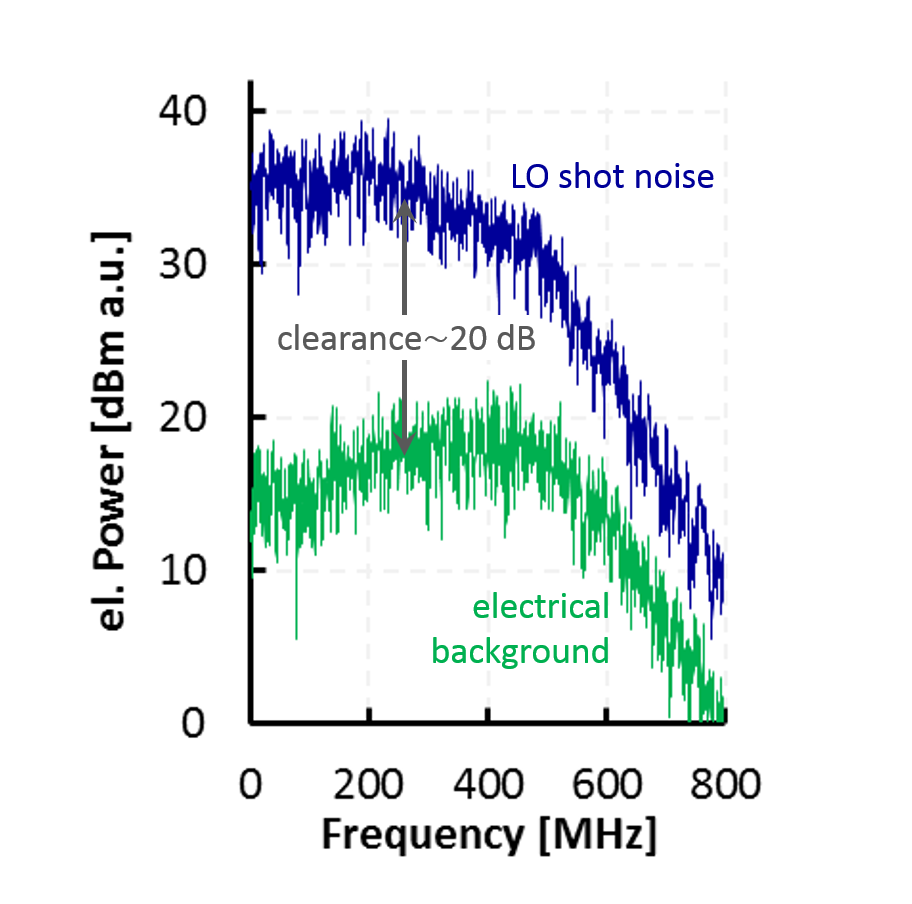}}
\subcaptionbox{}
	[0.6\linewidth]{\includegraphics[width=0.55\textwidth]{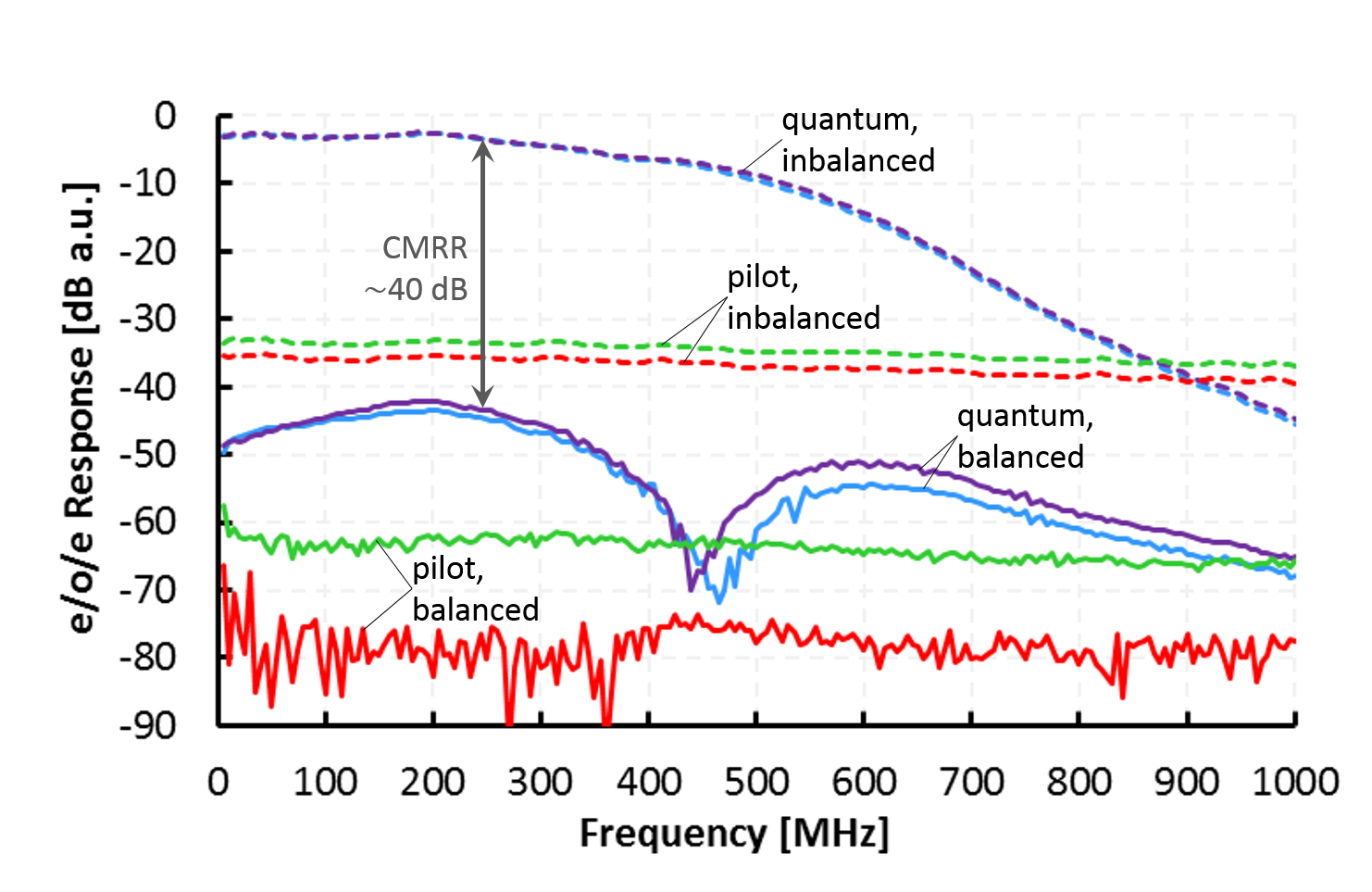}}
\caption{(a) Schematics of the receiver setup. An optical $\SI{90}{\degree}$-polarisation-diversity hybrid separates the incoming compound signal with respect to polarisation and mixes the quantum signal and pilot tone with the LO and routes them to the respective balanced receivers (one receiver for each basis in the quantum- and pilot branch). On the right side a screenshot of the oscilloscope traces illustrates the measurement data obtained by the four receivers: The QPSK quantum data with respect to time is depicted by the yellow ($I$) and green ($Q$) line, the periodic pilot tone by the smoother blue and red line. The bottom of the oscilloscope illustrates the pilot and quantum data in the frequency domain. (b) Shot-noise response and intrinsic electronic noise of the quantum receivers. The clearance is defined as the ratio of the two and indicates the receiver noise $\xi_{\text{det}}$ in shot-noise units. (c) CMRR of quantum- and pilot receivers. The lines correspond to the $I$- and $Q$ detectors of the quantum signal (purple and blue) and the pilot tone (green and red), respectively. At balanced response (solid lines), both PIN diodes of one detector were fed with the same optical power; to measure the imbalanced response (dashed lines), one of the two diodes was disconnected from the optical signal.}
\label{hybrid}
\end{figure*}

The receiver setup is illustrated in Fig.~\ref{hybrid}. An optically free-running local oscillator (LO, \emph{Teraxion PS-NLL}) with a narrow linewidth of $\Delta f< \SI{20}{kHz}$ and a total power of $\SI{12}{dBm}$ was used for coherent optical detection. Manual frequency alignment between $f_{T}$ and $f_{\text{LO}}$ was performed by current- and temperature-tuning in order to ensure an optical-frequency deviation much smaller than the symbol rate of the quantum data $|f_{T}-f_{\text{LO}}| \ll R_{\text{sym}}$.

Coherent intradyne reception of both tributaries was performed by use of a polarisation-diversity $90\text{\textdegree}$~hybrid (\emph{Kylia COH28-X}) which mixed the quantum signal and pilot tone with the local oscillator and routed them to their designated balanced detectors. The hybrid's insertion loss for the LO was $\SI{10}{dB}$ ($\SI{9}{dB}$ natural loss due to $1 \times 8$ splitting + $\SI{1}{dB}$ excess loss) -- at $\SI{12}{dBm}$ LO power corresponding to $\SI{2}{dBm}$ optical power per PIN diode (4 detectors/eight diodes). The insertion loss for the quantum signal and pilot tone amounted to $\SI{7}{dB}$ ($\SI{6}{dB}$ natural loss due to $1 \times 4$ splitting + $\SI{1}{dB}$ excess loss). The polarisation was manually realigned right before the optical hybrid using an in-line polarisation controller (\emph{Fiber Control FPC-1}, $\SI{0.7}{dB}$ insertion loss).

Two pairs of balanced detectors (one for each quadrature component of quantum signal and pilot) were used for opto-electronic signal conversion, each pair tailored to the specific needs of the respective signal tributary: The quantum data was detected using low-noise receivers (\emph{Insight BPD-1}) with a bandwidth of $\SI{360}{MHz}$, a clearance of $\SI{20}{dB}$ (Fig.~\ref{hybrid}(b)), common-mode rejection ratio (CMRR) of $\sim \SI{40}{dB}$ (Fig.~\ref{hybrid}(c)) and responsivity of $\SI{0.85}{A/W}$, corresponding to a quantum efficiency of $0.68$. Their low electronic noise (noise-equivalent power $< \SI{5}{pW/\sqrt{Hz}}$) is particularly important in case the receiver noise is attributed to the untrusted excess noise.

A set of high-bandwidth ($> \SI{1}{GHz}$) PIN/TIA receivers (\emph{Thorlabs PDB480C-AC}; Fig.~\ref{hybrid}(c)) was chosen for the stronger and therefore more robust pilot tone which requires a larger bandwidth and optical saturation limit but can, in return, tolerate more noise than measurements of the quantum signal since the pilot itself is not security-sensitive to eavesdropping attacks. Their responsivity was $\SI{0.95}{A/W}$, corresponding to a quantum efficiency of $0.76$.

After detection, the electrical I/Q signals were post-amplified, acquired in blocks of $2^{20}$ samples ($\approx \SI{1.05}{megasamples}$) by a real-time oscilloscope (\emph{Agilent Technologies DSO-X 91604A}) at a rate of 20~gigasamples per second (GS/s). At a symbol rate of $\SI{250}{Mbaud}$ this corresponds to 80~samples per symbol and a block size of $\approx 1.31\times 10^4$ symbols per block.\footnote{Although this oversampling allowed us to investigate various engineering aspects post-measurement, the sampling rate could in principle be reduced to the realm of commercially available ADCs ($\sim \SI{1}{GS/s}$).} Subsequently, the recorded data was fed to offline digital signal processing (DSP), as described in the next section.

\subsection{Digital Signal Processing}

\begin{figure*}
\centering
\includegraphics[width=0.9\linewidth]{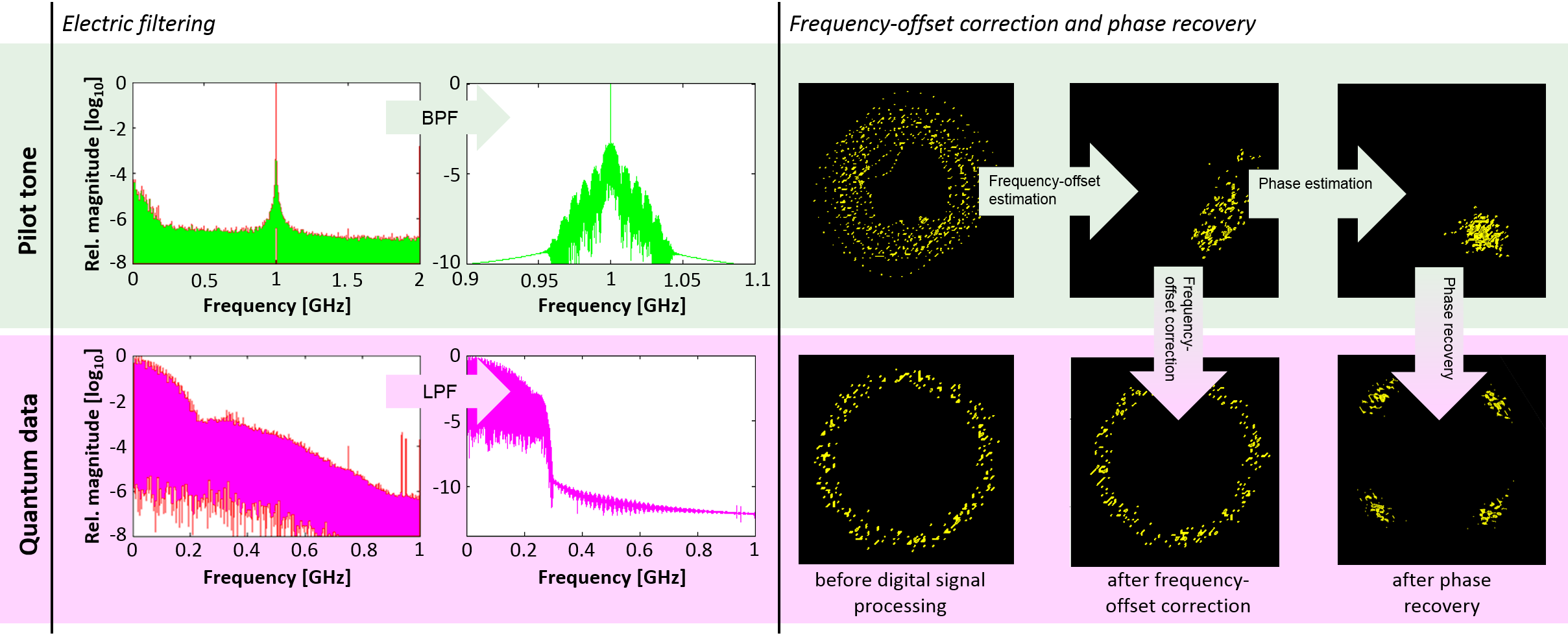}
\caption{Sequence of the digital signal processing (DSP). The pilot tone is bandpass-filtered (BPF) with a FWHM of $\SI{4}{MHz}$, the quantum data is lowpass-filtered (LPF) with $\SI{250}{MHz}$ cutoff. The phase-space constellations for the acquired pilot and the quantum data are shown before and after DSP. The optical-frequency offset $|f_{T}-f_{\text{LO}}|$ as well as the phase drift between LO and transmitter laser turns the 1-point pilot constellation and the QPSK constellation of the quantum signal into a ring. However, the optical phase of the pilot can be sampled by virtue of its high signal-to-noise ratio. In this way, the relative phase drift is estimated and applied to compensate the frequency offset and the phase drift of the quantum data. The original QPSK data can be recovered with good quality, as evidenced by the distinguishable constellation points.}
\label{DSP}
\end{figure*}

The individual DSP steps are illustrated in Fig.~\ref{DSP}. The first step consists of signal conditioning by means of spectral filtering of noise in the excess base- and pass-bandwidth. The quantum data was lowpass-filtered at $\SI{250}{MHz}$ (according to the symbol rate) and the pilot tone was bandpass-filtered at $\SI{4}{MHz}$, satisfying the trade-off between suppressing noise and at the same time covering the full frequency offset $|f_{T}-f_{\text{LO}}|$.

For the frequency-offset estimation the optical phase drift between the transmitted pilot and the LO was quantified by the rotation of the received pilot tone in phase space. Since the quantum signal was optically phase-locked to the pilot and had therefore experienced the same phase changes, the measured I/Q quadratures could be corrected using the rotation of the pilot tone which had been robustly acquired at high signal-to-noise ratio. More accurately, the frequency offset between LO and transmitter was determined by averaging the phase difference of subsequent pilot-tone measurements.

In order to remove remaining phase deviations due to the non-zero linewidth of the lasers, the measured quantum data points were first exponentiated by the power of four which aligned the four QPSK points -- spaced in multiples of $\pi/2$ -- to one angle in the complex plane up to phase deviations: $\exp^4(i\phi)=\exp^4(i[\phi+\pi/2])=\exp^4(i[\phi+\pi])=\exp^4(i[\phi+3\pi/2])$. Each symbol was then individually back-rotated and aligned to the mean angle.

The clock synchronisation was performed by measuring the correlation of the quantum data with the well-known PRBS7 sequence, or more precisely, by a convolution of the Fourier transform (FT) of the repeated PRBS7 sequence with a FT of the $I$ and $Q$ data string, respectively. (In case of true random modulation, synchronisation would require a time-interleaved preamble before each block of quantum data to accomplish periodic timing recovery.)

Finally, a CV-QKD parameter estimation was performed on the recovered quantum QPSK data to determine the excess noise and, therefore, the quality of the pilot scheme.

\section{Continuous-Variable Key-Transmission Performance} \label{sec_perf}

\begin{figure*}
\centering
\subcaptionbox{}
	[0.57\linewidth]{\includegraphics[width=0.56\textwidth]{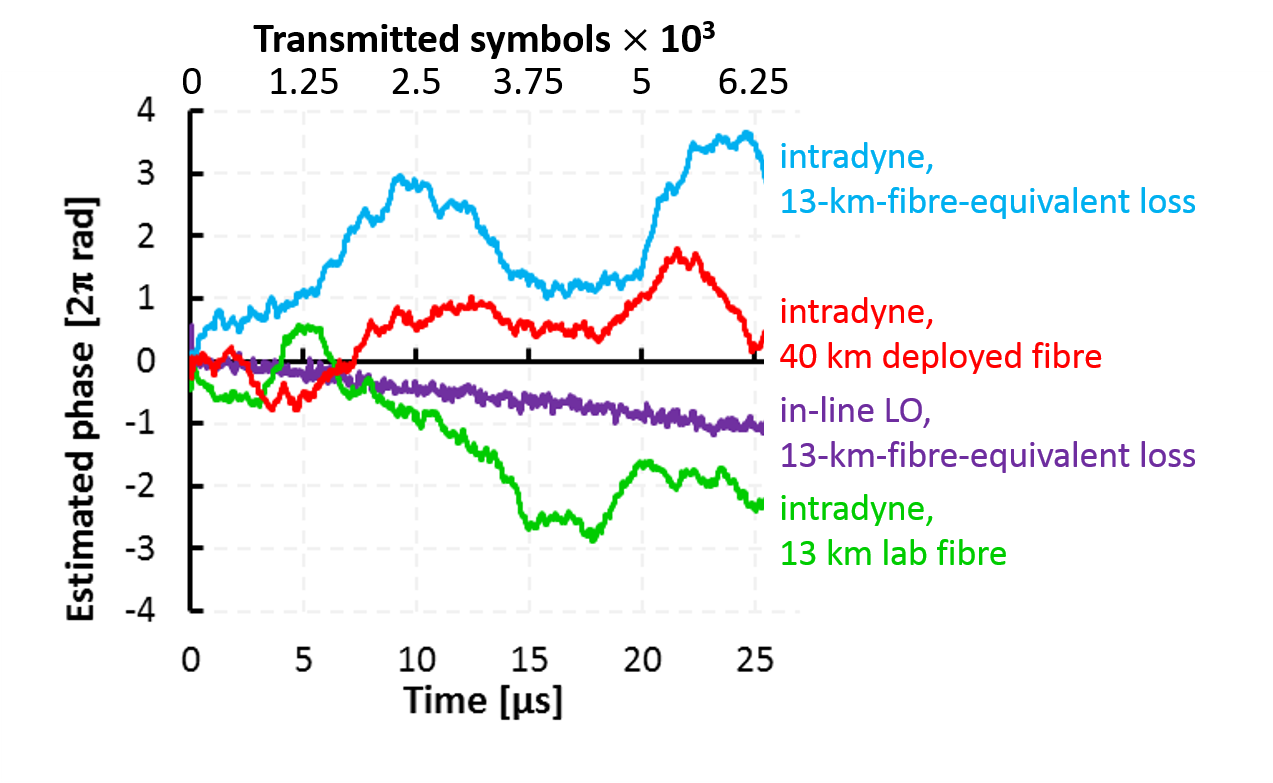}}
\subcaptionbox{}
	[0.33\linewidth]{\includegraphics[width=0.32\textwidth]{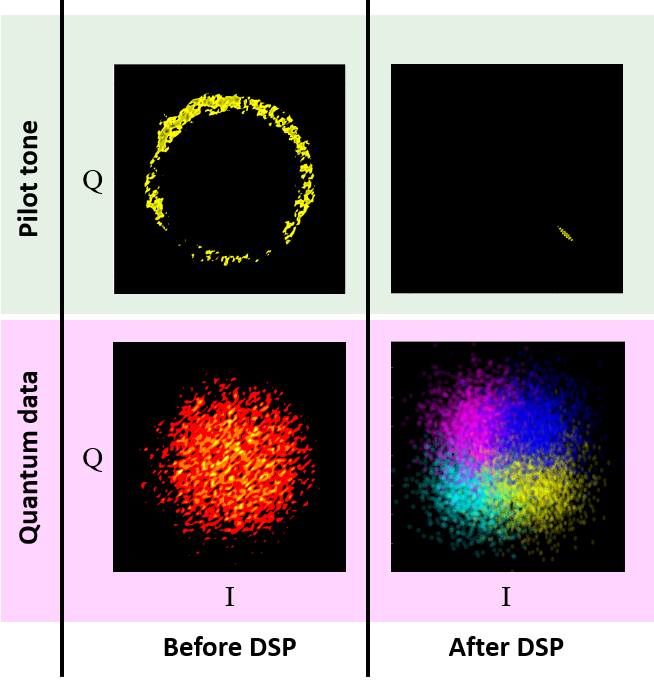}}
\caption{(a) Accumulated phase drifts over time at four different measurement scenarios. The secondary $x$-axis on the top displays the transmitted number of symbols corresponding to a symbol rate of $\SI{250}{Mbaud}$. The light-blue and purple line represent transmission over a short fibre where a channel loss corresponding to $\SI{13}{km}$ has been adjusted using a variable attenuator. (b) Phase-space representation of pilot- and quantum data at $\SI{13}{km}$ transmission distance and intradyne reception before and after digital-signal processing, i.e. frequency-offset correction and phase recovery.}
\label{phasespaceresults}
\end{figure*}

\begin{table*}
\centering
\begin{tabular}{|c|c|c|c|c|c|c|c|c|}
\hline 
\multicolumn{5}{|c|}{} & \multicolumn{2}{c|}{\textit{averaged calib.}} & \multicolumn{2}{c|}{\textit{worst-case calib.}} \\ 
\hline 
\textbf{Channel length [km]} & $\bm{T}$ & $\bm{V_{\text{mod}}}$ & $\bm{\braket{n_{B}}}$ & \textbf{SNR} & $\bm{\xi_{\text{tot}}}$ & $\bm{\xi_{\text{tot}}-\xi_{\text{det}}}$ &  $\bm{\xi_{\text{tot}}}$ & $\bm{\xi_{\text{tot}}-\xi_{\text{det}}}$ \\ 
\hline \hline 
1 & 0.60 & 3.0 & 0.92 & 0.90 & 0.036 & 0.015 & 0.067 & 0.044 \\ 
\hline 
4 & 0.53 & 4.1 & 1.1 & 1.06 & 0.023 & 0.0016 & 0.054 & 0.031 \\ 
\hline 
13 & 0.35 & 3.7 & 0.65 & 0.64 & 0.022 & 0.0010 & 0.053 & 0.030 \\ 
\hline 
40 & 0.10 & 12.5 & 0.63 & 0.62 & 0.026 & 0.0047 & 0.057 & 0.034 \\ 
\hline \hline
6 (no supp.\ carr.) & 0.50 & 6.4 & 1.5 & 1.44 & 0.079 & 0.057 & 0.11 & 0.090 \\ 
\hline 
13 (no supp.\ carr.) & 0.35 & 8.5 & 1.5 & 1.41 & 0.087 & 0.066 & 0.12 & 0.098 \\ 
\hline 
\end{tabular}
\caption{Measurement results after raw-key transmission over four channel lengths. The amount of excess noise that is attributed to the eavesdropper depends not only on whether one operates under the trusted-detector assumption ($\xi_{\text{Eve}}=\xi_{\text{tot}}-\xi_{\text{det}}$), but also on the way the calibration is performed: For the averaged calibration we used the mean of all the calibration measurements, for the worst-case calibration we only used the most disadvantageous one. The noise results indicate a strong advantage of optical carrier suppression over SSB without suppressed carrier (two bottom lines). Note that all excess-noise figures describe the quadrature variance in addition to the shot noise at the \emph{receiver} side and that the measured excess noise in each basis is $\xi/2$.}
\label{resulttable}
\end{table*}

The experiment was conducted over four different channel lengths using standard single-mode fibres of total length $\SI{1}{km}$, $\SI{4}{km}$, $\SI{13}{km}$ and $\SI{40}{km}$, respectively. The first three channels were realised by fibre spools in the lab, the latter one was provided by a $\SI{40}{km}$ deployed fibre in the city of Vienna.

The frequency offset amounted to $0$--$\SI{10}{MHz}$ and required readjustment of the LO frequency by current- and temperature-tuning in intervals of roughly 30 minutes. Figure~\ref{phasespaceresults}(a) illustrates the observed phase drifts at different transmission lengths and measurement scenarios, including an in-line-LO reception scheme that has been evaluated for the sake of comparison. For this purpose the optical carrier of the transmitter was reused as LO for coherent optical detection at the receiver side rather than using an independent, local LO. Although Fig.~\ref{phasespaceresults}(a) does confirm the expected phase robustness inherent to the in-line-LO scheme, we observed no performance advantage over our LLO intradyne scheme which exhibited peak phase-drift rates of less than $\SI{7}{rads/\mu s}$ -- at $\SI{250}{Mbaud}$ symbol rate corresponding to a maximum drift of only $\SI{28}{mrad}$ per symbol. These deviations could therefore be easily tracked and corrected post-measurement. A phase-space illustration of the results before and after digital signal processing in the case of $\SI{13}{km}$ intradyne reception is found in Fig.~\ref{phasespaceresults}(b).

In order to evaluate the performance of our pilot-tone concept, we performed the calibration of our measurements as well as the parameter estimation along the lines of Ref.~\cite{laudenbach2017continuous}. As primary indicator of the experimental quality we investigated the excess noise $\xi$, i.e.\ the quadrature variance in addition to the obligatory quantum shot noise. We define the excess noise referring to the \emph{receiver} (as opposed to the transmitter), such that the total quadrature variance at Bob is represented as

\begin{align}
V_{B}=V(\hat{I}_{B})=V_{B}(\hat{Q}_{B})=TV_{\text{mod}}+N_{0}+\xi,
\end{align}
where $T$ is the total channel transmittance, $V_{\text{mod}}$ is the modulation variance as applied by Alice (equal to twice the mean photon number per symbol) and $N_{0}$ is the quantum shot noise ($N_{0}=1$ in shot-noise units, SNU). (Note that in our notation the total excess noise comprises the electronic noise of the detectors, in literature often labelled as $\nu_{\text{el}}$.) Since we performed simultaneous measurement of $I$ and $Q$, we split the incoming signal into two arms, one for each quadrature basis. The variance in each arm is therefore

\begin{align}
V_{B}' = V(\hat{I}_{B}')=V(\hat{Q}_{B}')=\frac{T}{2} V_{\text{mod}}+N_{0}+ \frac{\xi}{2},
\end{align}
where we doubled the electronic detection noise of each basis ($\xi_{\text{det}}=2\xi'_{\text{det}}$) before integration into the total excess noise $\xi$ such that Eq. (2) holds. In contrast to the total variance above, the conditional variance $V_{B|A}$ relates Bob's measurement results to the symbols modulated by Alice (and therefore requires for Alice or Bob to disclose a certain fraction of their data). In general, it is represented as (here in the $I$-basis)

\begin{align}
V'_{B|A}=V \left( \sqrt{\frac{T}{2}}\hat{I}_{A}-\hat{I}_{B} \right).
\end{align}
In a discrete-modulation alphabet (like QPSK) $V'_{B|A}$ is simply the variance of all measurements that have been associated to one and the same symbol during the bit disclosure. Since the conditional variance in each basis is $V'_{B|A}=N_{0}+\xi/2 \stackrel{\text{SNU}}{=} 1+\xi/2$, the excess noise could be computed using

\begin{align}
\xi=2 \cdot (V'_{B|A}-1).
\end{align}
As another key parameter, the signal-to-noise ratio (SNR), was determined using

\begin{align}
\text{SNR} = \frac{\frac{T}{2} V_{\text{mod}}}{1+\frac{\xi}{2}} = \frac{V'_{B}}{V'_{B|A}}-1.
\end{align}

Table~\ref{resulttable} summarises the numeric experimental results over four distances using optical carrier-suppressed single-sideband modulation (oCS-SSB) of the pilot, processing $2^{20}$ samples ($\approx \SI{1.05}{megasamples}$) for each measurement.\footnote{Note that the numeric excess-noise results listed in Table~\ref{resulttable} refer to the receiver side and were multiplied by two in order to account for the balanced beamsplitter that splits the signal (and therefore also the excess-noise variance) in half for simultaneous $I$ and $Q$ measurement.}

\begin{figure}
\centering
\includegraphics[width=\linewidth]{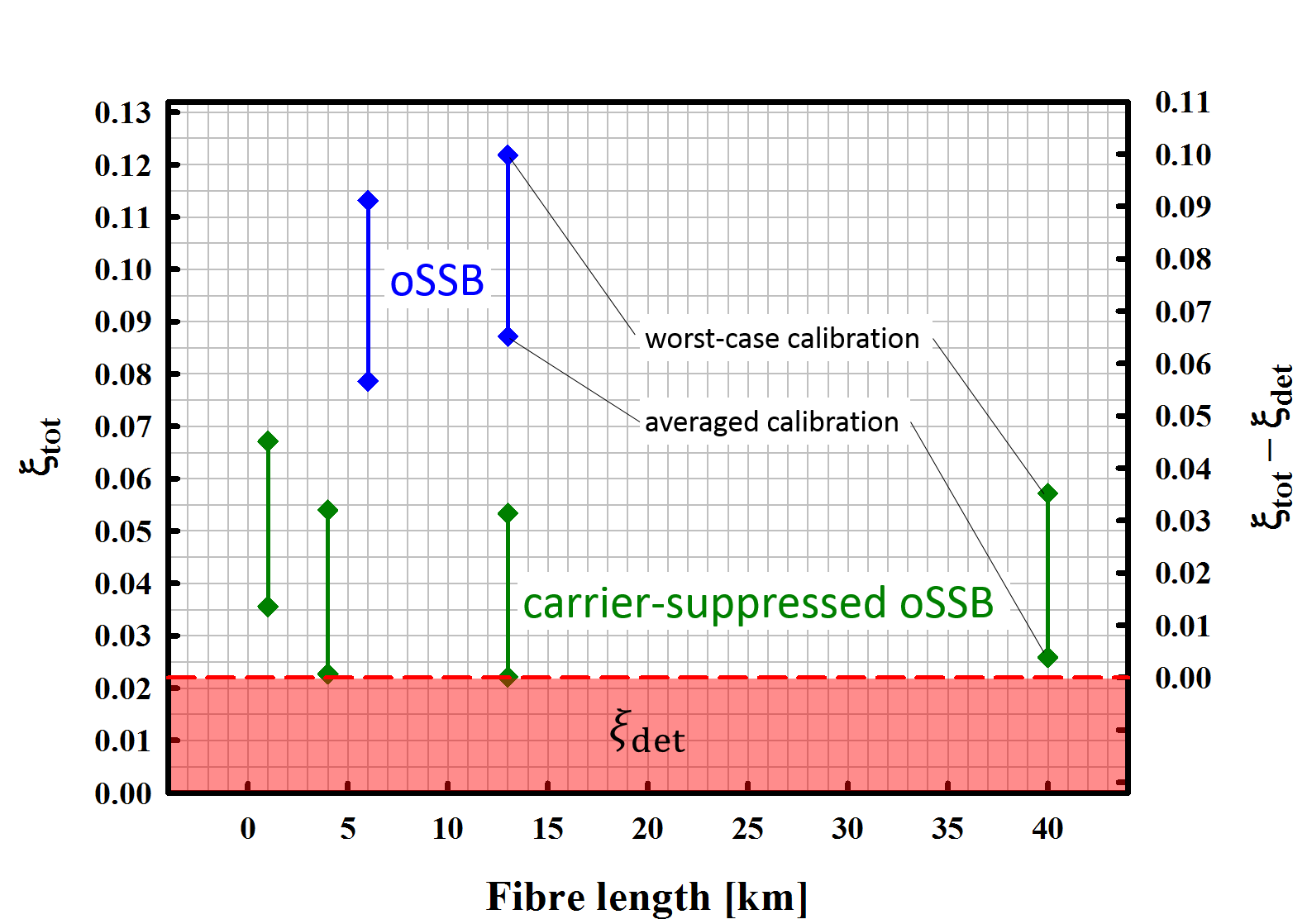}
\caption{Illustration of the measured excess noise at optical single-sideband modulation with suppressed optical carrier (green) and oSSB without carrier suppression (blue). The lower diamonds describe the noise results after averaged calibration, the upper ones represent the noise of the same measurement, but under worst-case calibration. The red area on the bottom represents the measured detection noise which can, under relaxed security assumptions, be regarded as trusted noise. In this case the noise attributed to an eavesdropper is $\xi_{\text{tot}}-\xi_{\text{det}}$ (right axis).}
\label{noiseresults}
\end{figure}

For calibration we performed a total of eight shot-noise measurements (standard deviation $\SI{1.2}{\%}$) and four measurements of the electronic noise (standard deviation $\SI{5.7}{\%}$), each of these measurements over a sample size of $2^{20}$.\footnote{We recently published a significantly improved method to reliably calibrate CV-QKD systems, see Ref.~\cite{brunner2019precise}.} Table~\ref{resulttable} lists the noise results for the case where the conversion was performed using the averaged calibration measurements as well as the results that we obtained when we, pessimistically, only used the one calibration measurement that yielded the highest conditional variance, and hence the largest excess noise $\xi$. The difference between the results for averaged and worst-case calibration can also be seen as a conservative upper bound for the uncertainty of the noise results, as indicated in Fig.~\ref{noiseresults}.

Moreover, for both calibration approaches, we list the total measured excess noise as well as the excess noise \emph{excluding} the intrinsic detection noise of the balanced receivers $\xi_{\text{det}}$. This is relevant for the relaxed assumption that the receivers in Bob's lab are regarded as trusted devices and therefore do not contribute to Eve's information~\cite{laudenbach2019analysis}.

As indicated by the table, the total measured excess noise $\xi_{\text{tot}}$ for the respective transmission distances amounted to values between $\SI{0.022}{SNU}$ and $\SI{0.036}{SNU}$ for averaged calibration and $\SI{0.053}{SNU}$ and $\SI{0.067}{SNU}$ for worst-case calibration. As for the excess noise excluding the detection noise $\xi_{\text{tot}}-\xi_{\text{det}}$, we obtained values between $\SI{0.0010}{SNU}$ and $\SI{0.015}{SNU}$ (averaged calibration), or respectively, between $\SI{0.030}{SNU}$ and $\SI{0.044}{SNU}$ (worst-case calibration). This compares beneficially to previous LLO implementations where the average excess noise excluding electronic receiver noise amounted to $\SI{0.0063}{SNU}$~\cite{qi2015generating}, $\SI{0.06}{SNU}$~\cite{soh2015self}, $\SI{0.015}{SNU}$~\cite{huang2015high}, $\SI{0.0022}{SNU}$~\cite{kleis2017continuous} and $\SI{0.0075}{SNU}$~\cite{wang2018high}, respectively (all values as measured by receiver).

The two bottom lines of Table~\ref{resulttable} depict our results using oSSB without suppressed carrier. Comparison of the results indicates an obvious advantage of optical carrier suppression yielding excess-noise figures which are lower by factors of $2\text{--}3$, as illustrated in Fig.~\ref{noiseresults}.

A direct experimental comparison of our intradyne LLO scheme with the inherently phase- and frequency-stable in-line-LO transmission yielded no noise penalty whatsoever being introduced by our method, therefore leaving no advantage to a transmitted LO other than, of course, a simpler experimental setup.

Our results indicate that the method of polarisation- and frequency multiplexing allows for higher symbol rates and, at the same time, lower excess noise compared to the currently more established time-multiplexing schemes~\cite{huang2015high, qi2015generating, soh2015self, wang2018pilot, wang2018high}. The method was demonstrated for QPSK modulation of the quantum signal but can, however, be implemented just as well for any modulation alphabet, including continuous Gaussian modulation for which the security analysis is understood best.

The estimated final secure-key rate depends on several additional parameters which are beyond scope of this local-LO demonstration: the reconciliation efficiency $\beta$, the frame-error rate $\text{FER}$, the fraction of the raw key $\nu$ that has to be disclosed during parameter estimation, the effects of finite blocklengths, and finally, the actual modulation alphabet that is used in the respective implementation of the scheme. To give an example, assuming $\beta=0.97$, $\text{FER}=0.05$, $\nu=0.25$, Gaussian modulation, coherent attacks, a trusted-receiver (detection noise not attributed to Eve) and averaged calibration, our experimental parameters obtained for the $\SI{40}{km}$ deployed fibre ($R_{\text{sym}}=\SI{250}{Mbaud}$, $V_{\text{mod}}=12.5$, $\text{SNR}=0.62$, $\xi_{\text{tot}}=0.026$, $\xi_{\text{tot}}-\xi_{\text{det}}=0.0047$) correspond to an asymptotic secure-key rate of $\SI{3.0}{Mbits/s}$ over $\SI{40}{km}$.

\section{Conclusion and Outlook}

We reported a reception method for CV-QKD with a free-running true local oscillator. We used a pilot tone of strong optical amplitude, multiplexed in frequency and polarisation to the quantum data, to establish the necessary phase reference. Experimental evaluation with a symbol rate of $\SI{250}{Mbaud}$ and over a transmission distance of up to $\SI{40}{km}$ yielded a low excess noise, confirming the robustness of the method. Not only closes our scheme the security loophole opened by a transmitted LO -- the approach of polarisation- and frequency multiplexing furthermore allows for higher symbol rates compared to time-multiplexing methods and for the deployment of optimised detectors for the quantum signal (low noise) and pilot (high bandwidth and saturation limit), respectively. Moreover, the cross-polarised preparation of signal and pilot tone as well as the suppression of the optical pilot carrier proved to be efficient methods to avoid crosstalk from the strong reference signal to the quantum channel. The raw-key transfer could be performed at a low-noise level, introducing no noise penalty with respect to the inherently phase-stabilised in-line-LO approach. Due to the demonstrated advantages, we believe that the proposed pilot-tone scheme is a promising candidate for any future implementation of high-performance CV-QKD transceivers.

As an outlook to our future work, we want to investigate and mitigate the reasons for the observed higher noise level at short transmission distance, implement our improved calibration procedure~\cite{brunner2019precise}, generate a secure key by CV-QKD post-processing under a QPSK security analysis and explore the co-propagation of classical data channels with the quantum channel.

\section*{Acknowledgements}

We acknowledge support from G. Humer, R. Lieger and E. Querasser for fruitful discussions and valuable technical support.

\end{document}